\documentclass[a4paper,11pt]{article}
\usepackage[utf8]{inputenc}
\usepackage{etex}
\usepackage{bm} 
\usepackage{graphicx}
\usepackage{epstopdf}
\usepackage{authblk} 
\usepackage{latexsym, amsmath, amsthm, amssymb}
\usepackage{parskip}
\usepackage{hyperref} 
\usepackage{color}
\usepackage{comment}
\usepackage{booktabs}
\usepackage[T1]{fontenc} 
\usepackage[english]{babel}
\usepackage[margin=0.8in]{geometry}
\usepackage{lscape}		
\usepackage{siunitx}	
\usepackage{natbib}
\usepackage[hyphenbreaks]{breakurl}

\def\ROI{{\rm ROI}}

\title{Pulse rate estimation using imaging photoplethysmography: generic framework and comparison of methods on a publicly available dataset}

\author[1,2,3]{Anton M. Unakafov\thanks{e-mail: anton@nld.ds.mpg.de}}

\affil[1]{Georg-Elias-Müller-Institute of Psychology, University of Goettingen, Goettingen, Germany}
\affil[2]{Theoretical Neurophysics Group, Max Planck Institute for Dynamics and Self-Organization, Goettingen, Germany}
\affil[3]{Leibniz ScienceCampus Primate Cognition, Goettingen, Germany}

\begin{document} 
\maketitle

\begin{abstract}
\noindent 
{\it Objective:} to establish an algorithmic framework and a benchmark dataset for comparing methods of pulse rate estimation using imaging photoplethysmography (iPPG). 
{\it Approach:} first we reveal essential steps of pulse rate estimation from facial video and review methods applied at each of the steps.   
Then we investigate performance of these methods for DEAP dataset \url{www.eecs.qmul.ac.uk/mmv/datasets/deap/} containing facial videos and reference contact photoplethysmograms. 
{\it Main results:} best assessment precision is achieved when pulse rate is estimated using continuous wavelet transform from iPPG extracted by the POS method (overall mean absolute error below 2 heart beats per minute). 
{\it Significance:} we provide a generic framework for theoretical comparison of methods for pulse rate estimation from iPPG and report results for the most popular methods on a publicly available dataset that can be used as a benchmark.   

\end{abstract}

\section{Introduction}\label{sect:Intro}

Heart rate (HR) is an important indicator of functional status, psycho-emotional state and health conditions in general.   
Traditionally HR is estimated from electrocardiogram or photoplethysmogram (PPG); however, both techniques require contact sensors, which can be disadvantageous \citep{NASA2011}, whereas non-contact HR estimation is useful, for example, for detecting driver drowsiness or abnormal state \citep{SahayadhasSundarajMurugappan2012}. 
 
For a non-contact assessment of pulse rate (equivalent of HR obtained from indirect peripheral measurements) imaging photoplethysmogramm (iPPG) analysis has been proposed \citep{TakanoOhta2007,VerkruysseSvaasandNelson2008}.  
Similarly to contact PPG, iPPG is acquired by measuring variations in the intensity of light reflected by the skin 
(see \cite{Allen2007, TamuraMaedaSekineYoshida2014} for details), but a video camera is used instead of simple photodetector. 
Then iPPG is computed from sequence of images, usually acquired from face or palm. 
Theoretical underpinnings of imaging photoplethysmography are provided in \cite{Huelsbusch2008, kamshilin2015newLook, WangDenBrinkerStuijkDeHaan2017}. 

Rapid development of iPPG analysis \citep{Tarassenko2014AR, Mcduff2015survey, SunThakor2016} emphasizes importance of comparing various algorithms for iPPG-based pulse rate estimation. 
Theoretical comparison is complicated since algorithms for iPPG acquisition consist of multiple steps that are often non-uniformly described. 
For empirical comparison a publicly available benchmark dataset is required since 
pulse rate estimates reported in different studies are not comparable due to the differences in
experimental conditions. 
However, to the best of our knowledge no suitable dataset has been proposed for iPPG benchmarking\footnote{Uncompressed video normally used for iPPG acquisition is too large for publishing on-line. An attempt to compare existing algorithms on a publicly available MAHNOB dataset \citep{MAHNOB2013} was made in \citep{LiChenZhaoPietikainen2014}.
However, this dataset seems to be unsuitable for iPPG benchmarking as the videos underwent strong compression making consistent iPPG extraction impossible \citep{WangDenBrinkerStuijkDeHaan2017}.}.  

To overcome the problems of comparing algorithms we suggest a generic algorithmic framework describing main steps of iPPG-based pulse rate estimation; we discuss popular methods employed at various steps and compare their performance on a publicly available dataset \citep{DEAP2012}, containing facial video and reference contact PPG.
We report experimental results demonstrating how the choice of the methods for each step influences overall quality of pulse rate estimation. 

Our framework consists of five steps\footnote{A similar three-step framework was proposed in \citep{RouastAdamChiongCornforthLux2016}, but that article gives an overview of iPPG acquisition while here we focus on the algorithmic details of iPPG processing steps.}.
Methods used at single steps of pulse rate estimation were previously compared in \citep{HoltonEtAl2013, CuiFuHongZhangShu2015, WangDenBrinkerStuijkDeHaan2017}; here we combine methods used at each of five steps to find their optimal configurations. 


\section{Materials and Methods}\label{sec:Methods}

\subsection{Dataset Description}\label{subsec:Dataset}
The Dataset for Emotion Analysis using EEG, Physiological and video signals (DEAP, \cite{DEAP2012}) contains physiological recordings and frontal face videos of 22 human volunteers watching music videos in 40 one-minute trials.  
We denote trials as P$x$ T$y$, where $x$ is the number of participant in DEAP and $y$ is the number of trial. 
Altogether, DEAP dataset consists of 861 one-minute trials with facial video and reference contact PPG data (37 trials for P11; 39 for P3, P5 and P14; 40 for other participants).    
We reject 13 trials where large part of the face was occluded (P4 T17; P6 T24; P12 T14, T18; P15 T12, T16, T23; P18 T4, T10; P22 T13, T18-T20) since for these videos stable iPPG acquisition was impossible.		

Videos were recorded in DV PAL format using a SONY DCR-HC27E camcorder and transcoded to 50 FPS deinterlaced video using the h264 codec. 
The resolution of all the videos is $720\times586$.

Contact PPG was acquired from the left thumb. 
We computed reference pulse rate values from PPG by determining intervals between diastolic minima \citep{SchaeferVagedes2013} using the method proposed in \cite{Elgendi2013systolic}\footnote{When applying this method to contact PPG from DEAP, we realize that it does not detect some diastolic minima since their amplitudes vary significantly. 
To alleviate this problem we introduce two modifications. 
First, to detect minima with varying amplitudes we determine offset level $\alpha$ \cite[Eq (7)]{Elgendi2013systolic} not as mean of the whole signal, but as running mean over window of 7~s.  
Second, to reject false positives that may arise from the first modification we add a post-processing step:  
after method detects diastolic minima $\rm{DM}_1, \rm{DM}_2,\ldots,\rm{DM}_N$, we reject $\rm{DM}_i$ if it holds 
\begin{equation*}
	\rm{DM}_{i+1} - \rm{DM}_{i-1} < \frac{2}{3}\min\Big(\rm{DM}_{i-1} - \rm{DM}_{i-3},  \frac{2.3}{N-1}\sum\limits_{j=2}^{N}(\rm{DM}_{j} - \rm{DM}_{j-1})\Big),
\end{equation*}  
where coefficients $\frac{2}{3}$ and $2.3$ are selected empirically.
}.


\subsection{Methods}\label{subsec:Methods}
\begin{figure}[!htbp]
	\centering
	\includegraphics[scale=0.032]{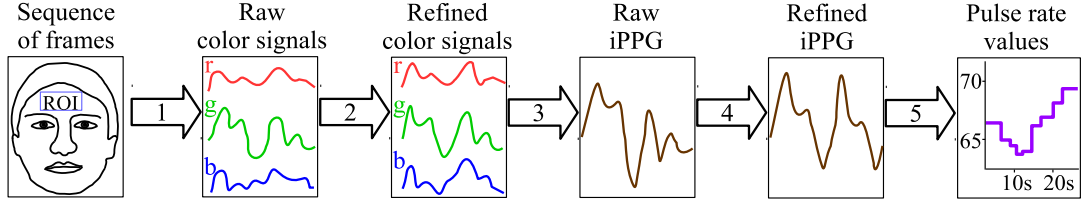}
	\caption{Five steps of pulse rate estimation from facial video using iPPG.}
 	\label{fig:scheme}
\end{figure}

In this section we propose a generic algorithmic framework of iPPG-based pulse rate estimation. 
It takes as an input a sequence of $T$ RGB frames; $t$-th frame for $t = 1,2,\ldots,T$ consists of pixels given by vectors $\bm{c}_{i,j}(t) = \big(r_{i,j}(t),g_{i,j}(t),b_{i,j}(t)\big)^\intercal$, where $r_{i,j}(t)$, $g_{i,j}(t)$, $b_{i,j}(t)$ are the red, green and blue channels for the pixel with coordinates $(i,j)$; $\bm{v}^\intercal$ stands for the transposed vector $\bm{v}$.
The algorithm consists of five steps schematically shown in Fig.~\ref{fig:scheme}; below we consider them in details.

\begin{figure}[ht]
	\centering
	\includegraphics[scale=0.032]{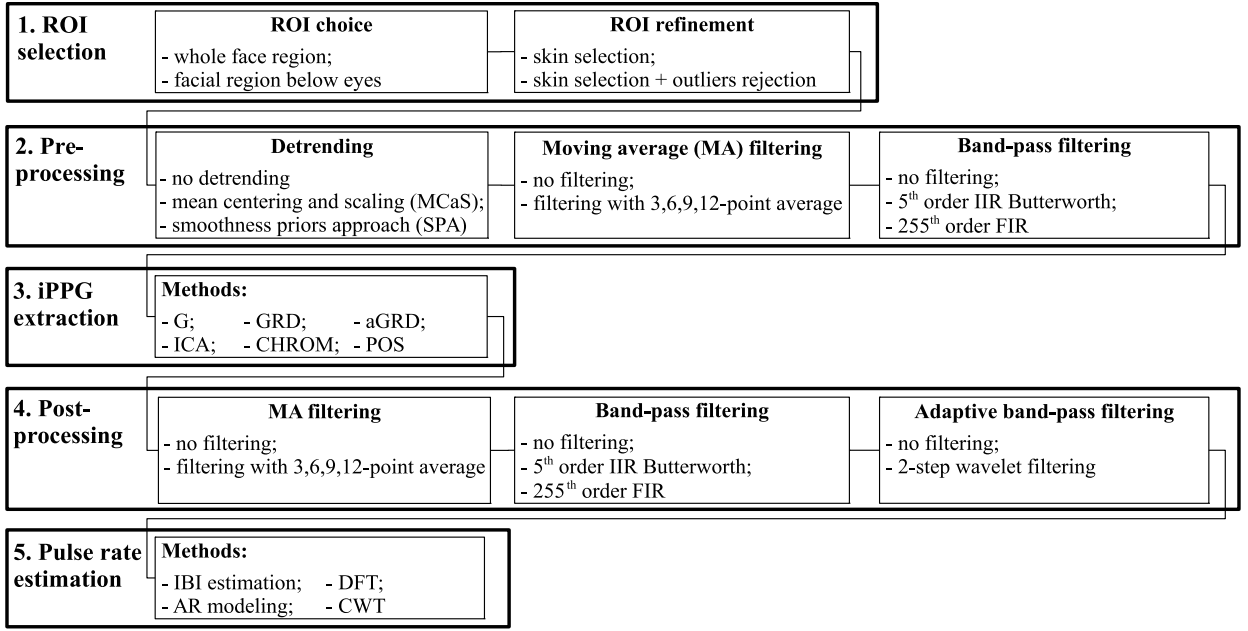}
	\caption{Scheme of considered methods. 
	Big blocks represent five steps of pulse rate estimation (see Figure~\ref{fig:scheme}), each box inside the block represents a sub-step and contains a list of methods used at this sub-step. We try various combinations of methods, each time taking one method for every sub-step.}
 	\label{fig:testScheme}
\end{figure}

\begin{enumerate}
	\item For every frame $t =1,2,\ldots,T$ select the region of interest $\ROI(t)$ as a set of pixels containing PPG-related information, and compute average color intensities over ROI (color signals):
	\begin{equation}
		\bm{c}^0(t) = \big(r^0(t),g^0(t),b^0(t)\big)^\intercal = \frac{1}{\vert \ROI(t) \vert}\sum_{(i,j) \in \ROI(t)} \bm{c}_{i,j}(t),		
	\end{equation}\label{eq:roi_averaging}	
	where $\vert \ROI(t) \vert$ is the number of pixels in $\ROI(t)$ (see Subsection~\ref{subsec:ROI} for $\ROI(t)$ selection).	
	\item Compute refined color signals $\bm{c}(t) = \big(r(t),g(t),b(t)\big)^\intercal$ by pre-processing $\bm{c}^0(t)$ (Subsection~\ref{subsec:Preproc}). 	
	\item Extract raw iPPG as a combination of refined color signals:
	\begin{equation*}
		\mathrm{iPPG}^{0}(t) = w_\mathrm{r}(t)r(t) + w_\mathrm{g}(t)g(t) + w_\mathrm{b}(t)b(t)
	\end{equation*}		
	with weights $w_\mathrm{r}(t),w_\mathrm{g}(t),w_\mathrm{b}(t) \in \mathbb{R}$ (see Subsection~\ref{subsec:IppgEstimation} for weights calculation).
	\item Post-process raw  signal $\mathrm{iPPG}^{0}(t)$ to get refined signal $\mathrm{iPPG}(t)$ (Subsection~\ref{subsec:Postproc}).	
	\item Estimate pulse rates from processed iPPG signal (Subsection~\ref{subsec:PulseRate}).

\end{enumerate}
We test several popular methods for every step of estimation algorithm (Figure~\ref{fig:testScheme}) in order to find out which combinations of methods provide most precise pulse rate estimation. 

\subsubsection{Selecting Region of Interest}\label{subsec:ROI}

To compute color signals $\bm{c}^0(t)$ by \eqref{eq:roi_averaging}, color intensities $\bm{c}_{i,j}(t)$ are averaged over ROI. 
As PPG-induced variations of facial color are weak in comparison with noise and artifacts,  
the aim of ROI selection is to choose pixels containing maximal pulsatile information, so that averaging reduces noise while preserving the iPPG signal.
ROI selection consists of two sub-steps: initial choice of facial region for iPPG acquisition ({\bf ROI choice}) and excluding irrelevant pixels ({\bf ROI refinement}). 

{\bf ROI choice.} The most popular approach is to take ROI as a rectangle encompassing the whole-face region \citep{LewandowskaRuminskiKocejkoNowak2011, PohMcDuffPicard2011, deHaanJeanne2013, MannapperumaEtAl2014ICAlimits}.  
	Other popular regions are the whole face excluding eye region \citep{McduffGontarekPicard2014remote, LiChenZhaoPietikainen2014} and forehead \citep{VerkruysseSvaasandNelson2008}.   	
	
	In DEAP dataset for some participants EEG cap covers most of the forehead, which hinders using the forehead region; therefore we consider the whole-face region and the facial region below eyes. 
	In both cases we detect facial rectangle for each frame by the commonly-used cascade classifier \citep{CascadeClassifier2017} constructed by means of the Viola-Jones algorithm \citep{ViolaJones2001}.
	We take the width of ROI equal to 80\% of the estimated face width as recommended in \citep{PohMcDuffPicard2011}.

{\bf ROI refinement.} Even when ROI is selected properly, some pixels may not contain iPPG signal. 
Examples include non-skin pixels (for instance, hair), over or under-lit areas, damaged pixels in the sensor. 
To exclude such pixels ROI-refinement methods are used, here we consider two of them.

First, non-skin pixels are discarded. 
This is an essential part of ROI refinement for DEAP since in many videos cables hang in front of participants' faces.  	
We use simple HSV masking\footnote{HSV color model is generally considered to be most useful for skin detection \citep{ZaritSuperQuek1999,VezhnevetsSazonovAndreeva2003}, another popular choice is YCbCr model \citep{BousefsafMaaouiPruski2013}. See also \citep{WangStuijkDeHaan2015skinSelection} for a more elaborate approach to skin selection for iPPG acquisition.}: pixels with hue, saturation or value outside of the ranges $[0^\circ, 46^\circ]$, $[23, 132]$ and $[88, 255]$, respectively, are considered non-skin and discarded (ranges are selected empirically as providing effective skin selection for the entire dataset).
	
Then we reject pixels that differ considerably from other pixels in ROI (outliers). 
Namely, we discard pixels $(i,j)$ that do not satisfy the following inequality \citep{TasliGudiUyl2014}:
		\begin{equation*}
			\vert\bm{c}^0(t)- \bm{c}_{i,j}(t)\vert < \gamma \bm{\sigma}_\ROI(t), \;\text{ where } 
			\bm{\sigma}_\ROI(t) = \sqrt{\frac{1}{\vert \ROI(t) \vert} \sum _{(i,j) \in \ROI(t)}\big(\bm{c}_{i,j}(t) - \bm{c}^0(t)\big)^2}.
		\end{equation*}
In \citep{TasliGudiUyl2014} $\gamma = 3$ is used; since this value does not provide effective outliers rejection for DEAP videos,  we take $\gamma = 1.5$.

Another important part of ROI refinement is motion compensation \citep{KumarVeeraraghavanSabharwal2015,WangStuijkDeHaan2015skinSelection}. 
Here we do not use it since there is no prominent head movements in videos from DEAP dataset.



\subsubsection{Pre-processing of Color Signals}\label{subsec:Preproc}

At this step refined color signals $\bm{c}(t)$ are computed from raw signals $\bm{c}^0(t)$  for $t = 1,\ldots,T$ by suppressing noise and artefacts.
To preserve relevant information, frequency components in human heart rate bandwidth (40--240 beats per minute (BPM), which corresponds to 0.65--4~Hz) should not be suppressed.  
Typical pre-processing  sub-steps are detrending, band-pass and moving average filtering (see Figure~\ref{fig:testScheme}, Step 2). They are often used in combination \citep{HoltonEtAl2013, LiChenZhaoPietikainen2014}, but some sub-steps can be omitted or applied at post-processing (Step 4, see Subsection~\ref{subsec:Postproc}). 

{\bf Detrending} is important since pulsatile component of iPPG has much lower amplitude than the slowly-varying baseline \citep{Huelsbusch2008}. 
A simple detrending method consists in mean-centering and scaling the signal (MCaS, \cite{deHaanJeanne2013}): 
	\begin{equation*}
		\bm{c}(t) = \frac{\bm{c}^0(t) - \bm{m}(t, L)}{\bm{m}(t, L)},
	\end{equation*}	
where $\bm{m}(t, L) = \frac{1}{L}\sum\limits_{k=0}^{L-1}\bm{c}(t-k)$ is an $L$-point running mean of color vectors $\bm{c}(t)$; we take $L$ corresponding to 1~s.
Using MCaS is required for many methods of iPPG extraction (see Subsection~\ref{subsec:IppgEstimation}). 

Another popular detrending method is smoothness priors approach (SPA, \cite{TarvainenRantaAhoKarjalainen2002}) used in \citep{PohMcDuffPicard2011, LiChenZhaoPietikainen2014}. To remove trend without affecting the heart rate bandwidth we employ SPA with control parameter $\lambda = 300$, which suppresses frequencies below 0.55~Hz, see \citep{TarvainenRantaAhoKarjalainen2002} for details.

{\bf Moving average} (MA) filtering smooths the signal and suppresses high-frequency noise. MA filtering with $M$-point average is provided by the following equation:
	\begin{equation*}
		\bm{c}(t) = \frac{1}{M}\sum_{k=0}^{M-1}\bm{c}^0(t-k).
	\end{equation*}		
When choosing $M$ one should take into account that $M$-point MA filter suppresses frequencies $\frac{n}{M} F_\text{SR}$ for $n =1,2,\ldots$, where $F_\text{SR}$ is the sampling rate of the signal, see \cite[Chapter~16]{Smith1997DSP} for details. 
	Since human pulse rate can reach 4 Hz, we recommend using $M<\frac{1}{4}F_\text{SR}$. For instance, $F_\text{SR} = 50$ Hz requires $M \leq 12$, thus we consider MA filtering with 3-, 6-, 9- or 12-point average. 
	

{\bf Band-pass} filtering suppresses frequency components outside the heart rate bandwidth. 
Here we employ two commonly used filters, either the 255-th order finite impulse response (FIR) filter with linear phase designed using the Hamming window \citep{LewandowskaRuminskiKocejkoNowak2011,PohMcDuffPicard2011, LiChenZhaoPietikainen2014}	or the 5th order Butterworth infinite impulse response (IIR) filter \citep{Sun2013noncontact}.


\subsubsection{Extracting Photoplethysmogram from Color Signals}\label{subsec:IppgEstimation}

This step (Figure~\ref{fig:testScheme}, Step 3) can be represented as:
	\begin{equation*}
		\mathrm{iPPG}^{0}(t) = \bm{w}(t) \cdot \bm{c}(t) = w_\mathrm{r}(t)r(t) + w_\mathrm{g}(t)g(t) + w_\mathrm{b}(t)b(t),
	\end{equation*}
	where $\bm{w}(t) = \big(w_\mathrm{r}(t),w_\mathrm{g}(t),w_\mathrm{b}(t)\big)^\intercal  \in \mathbb{R}^3$ are weights of color signals. For computing these weights the following methods are often used:
	\begin{itemize}
		\item Estimating iPPG by the green signal ({\bf G} method). 
		This approach is popular \citep{Tarassenko2014AR, CuiFuHongZhangShu2015} due to its simplicity, 
		in this case $\bm{w}(t) = (0,1,0)$ that is 
		\begin{equation*}
			\mathrm{iPPG}^{0}(t) = g(t).
		\end{equation*}

		\item Estimating iPPG by the green signal while the red signal is considered as containing artefacts only (green-red difference or {\bf GRD} method).
		 Here $\bm{w}(t) = (-1,1,0)^\intercal$, thus 
		\begin{equation*}
			\mathrm{iPPG}^{0}(t) = g(t) - r(t).
		\end{equation*}		 
		This method was first proposed in \cite[Chapter~6]{Huelsbusch2008} as a robust alternative to G method. 

		\item Adaptive green-red difference ({\bf aGRD}, \cite{FengPoXuLiMa2015}) computes iPPG as 
		\begin{equation}\label{eq:aGRD}
			\mathrm{iPPG}^{0}(t) = \Vert\bm{c}^0(t)\Vert\Big(\frac{g(t)}{g^0(t)} - \frac{r(t)}{r^0(t)}\Big),
		\end{equation}		 
		 where $\Vert\bm{c}^0(t)\Vert = \sqrt{\big(r^0(t)\big)^2+\big(g^0(t)\big)^2+\big(b^0(t)\big)^2}$. 		
		Pre-processing is essential for this method since otherwise $g(t)\equiv g^0(t)$ and $r(t)\equiv r^0(t)$ result in $\mathrm{iPPG}^{0}(t) \equiv 0$. Originally, a band-pass filtering is used \citep{FengPoXuLiMa2015}.
		
		\item Decomposing color signals into components by means of blind source separation (BSS) and choosing the component with the most prominent peak in the heart rate bandwidth. 
		Independent component analysis ({\bf ICA}) is the most popular BSS technique for iPPG computation \citep{HoltonEtAl2013, WangDenBrinkerStuijkDeHaan2017}.  
		Here we use JADE algorithm of ICA by \cite{Cardoso1999} as suggested in \citep{PohMcDuffPicard2010, PohMcDuffPicard2011}\footnote{ICA-based iPPG extraction incorporates specific pre-processing (subtracting the mean and dividing by standard deviation of each channel \citep{McduffGontarekPicard2014OCG}; see \citep{deHaanLeest2014} for the criticism of this approach) and post-processing (inverting iPPG if it was flipped during ICA \citep{McduffGontarekPicard2014OCG}). We use these methods as a part of ICA but do not describe them separately due to their limited applicability for non-ICA-based iPPG extraction.}.

		
		\item {\bf CHROM} method \citep{deHaanJeanne2013} employs a model of PPG-induced variations in color intensity and defines iPPG signal as
		\begin{equation}
			\mathrm{iPPG}^{0}(t) = x_1(t) - \frac{\sigma_1(t,L)}{\sigma_2(t,L)}x_2(t),
		\end{equation} 		
		where $\sigma_1(t,L)$, $\sigma_2(t,L)$ are $L$-point running standard deviations of $x_1(t) = 0.77r(t)-0.51g(t)$ and $x_2(t) = 0.77r(t)+0.51g(t)-0.77b(t)$, respectively: 			
		\begin{equation}\label{eq:CHROM_sigma}
			 \sigma_i(t,L) = \sqrt{\frac{1}{L-1}\sum_{k=0}^{L-1}x_{i}(t-k)^{2}-
			 					   \frac{1}{L(L-1)}\left(\sum_{k=0}^{L-1}x_{i}(t-k)\right)^{2}}
		\end{equation}
		for $i =1,2$. We follow \citep{deHaanJeanne2013} in taking $L$ corresponding to 1.6 s.

		\item The recently proposed {\bf POS} method \citep{WangDenBrinkerStuijkDeHaan2017} can be considered as an improved and simplified version of CHROM:
		\begin{equation*}
			\mathrm{iPPG}^{0}(t) = x_1(t)+\frac{\sigma_1(t,L)}{\sigma_2(t,L)}x_2(t), 
		\end{equation*} 			
		where $\sigma_1(t,L)$ and $\sigma_2(t,L)$ are $L$-point running standard deviations \eqref{eq:CHROM_sigma} of $x_1(t) = g(t)-b(t)$ and $x_2(t) = g(t)+b(t)-2r(t)$, respectively. We take $L$ corresponding to 1.6 s as suggested in  \citep{WangDenBrinkerStuijkDeHaan2017}.
	\end{itemize}	

In order to make CHROM and POS compliant with our generic framework, we introduce minor algorithmic changes not affecting the nature of the methods. 
Namely, we use running means and standard deviations instead of computing iPPG signal in overlapped windows. 
For the considered dataset the performance of our modified versions is slightly better than of the original methods.  

Note that special pre-processing is required for some iPPG extraction methods, namely MCaS detrending for GRD, CHROM and POS and band-pass filtering for aGRD. When testing the effect of pre-processing (see Figure~\ref{fig:testScheme}, Step 2) we always use required pre-processing with these iPPG extraction methods.

\subsubsection{Post-processing of Imaging Photoplethismogram}\label{subsec:Postproc}
Post-processing (Figure~\ref{fig:testScheme}, Step 4) improves quality of iPPG signal and is especially necessary if noise and artifacts were not removed at pre-processing (Step 2) or if iPPG was extracted at Step 3 in a non-linear fashion (which is the case for aGRD, ICA, CHROM and POS). 
Here we consider three typical sub-steps of post-processing: {\bf band-pass}, {\bf MA} and {\bf adaptive band-pass (ABP) filtering}.  

{\bf Band-pass} and {\bf MA filtering} described in Subsection~\ref{subsec:Preproc} as pre-processing sub-steps can be also used at post-processing \citep{PohMcDuffPicard2010, PohMcDuffPicard2011}, this results in different iPPG signal for all considered methods of iPPG extraction except for linear G and GRD methods. 

{\bf ABP filtering} assumes that frequency components of iPPG signal pertaining to pulse rate have relatively high power; then weak components correspond to noise and should be suppressed \citep{Huelsbusch2008, BousefsafMaaouiPruski2013, WangStuijkDeHaan2015skinSelection, FengPoXuLiMa2015}. 
Here we use a two-step {\bf wavelet filtering} suggested in \citep{BousefsafMaaouiPruski2016}\footnote{First we perform continuous wavelet transform of iPPG and filter wavelet coefficients with a wide Gaussian window centered at scale corresponding to the maximum of squared wavelet coefficients averaged over 15~s temporal running window. Then we apply usual Gaussian filter. The filtered signal is reconstructed by performing the inverse continuous wavelet transform. See \cite{BousefsafMaaouiPruski2016} for details.}.  
 		
\begin{figure}[!htbp]
	\centering
	\includegraphics[scale=0.55]{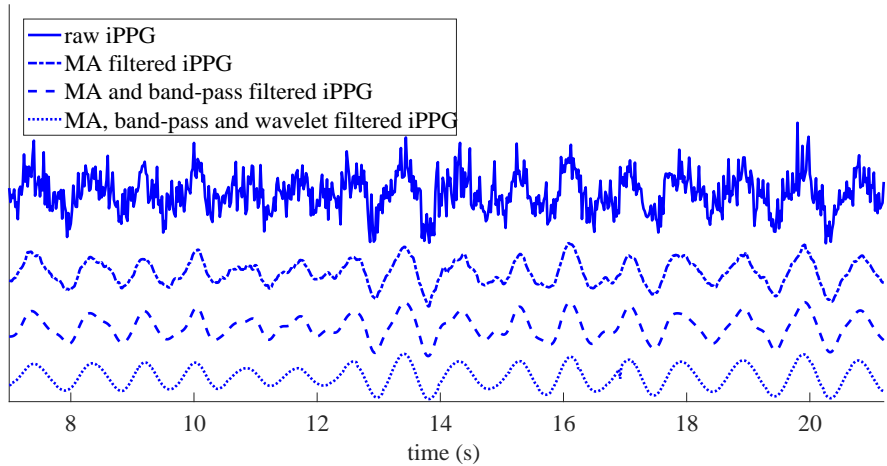}
	\caption{Effect of various post-processing methods on iPPG extracted from P1 T24 data using POS.} 
	
 	\label{fig:PostprocessingEffectOnIPPG}
\end{figure}

Modifications of iPPG signal provided by MA, band-pass and wavelet filtering are shown in Figure~\ref{fig:PostprocessingEffectOnIPPG}. 

\subsubsection{Estimation of Pulse Rate}\label{subsec:PulseRate}

We consider here four most popular methods of pulse rate estimation (Figure~\ref{fig:testScheme}, Step 5).

\begin{itemize}
\item {\bf Interbeat interval (IBI)} estimation is the most direct way to assess pulse rate, however this approach is rarely used for iPPG since precise IBI estimation is often problematic \citep{SchaeferVagedes2013, Elgendi2013systolic, Kamshilin2016accurate}. 
IBI corresponds to a cardiac cycle; thus momentary pulse rate is equal to the inverse IBI duration. 
IBI is usually defined for iPPG as time between successive systolic peaks \citep{SchaeferVagedes2013} using some method of peak detection; here we employ method from \cite{Elgendi2013systolic} with modifications described in Subsection~\ref{subsec:Dataset}, see Figure~\ref{fig:PPGsignal} for an illustration. 
For accurate IBI estimation we increase sampling rate of iPPG signal from 50 to 250~Hz using cubic spline interpolation as suggested in \citep{TakanoOhta2007}.		

\begin{figure}[!htbp]
	\centering
	\includegraphics[scale=0.55]{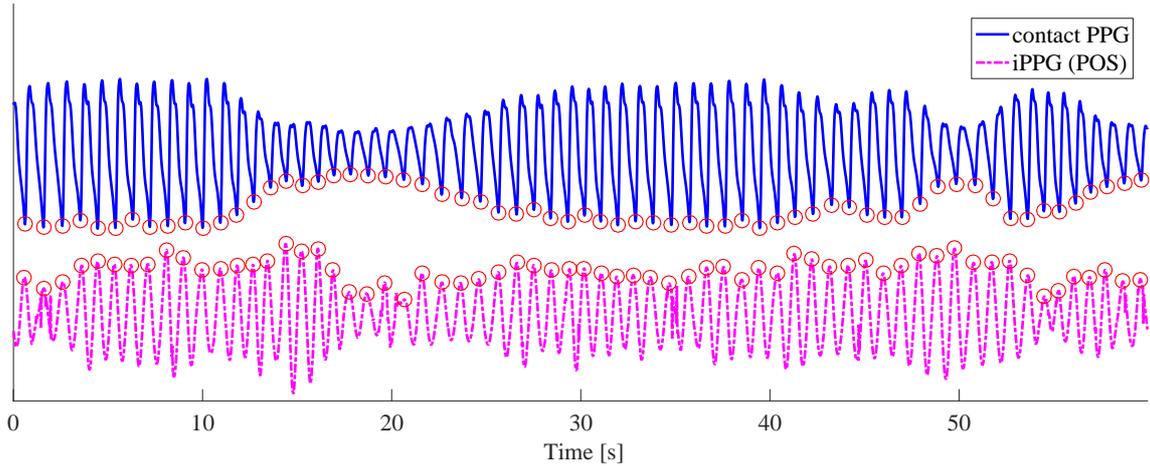}
	\caption{Contact PPG and iPPG signal (extracted using POS and post-processed by MA, band-pass and wavelet filtering) for P1 T24, red circles indicate diastolic minima for PPG and systolic peaks for iPPG detected using algorithm from \cite{Elgendi2013systolic} with modifications described in Subsection~\ref{subsec:Dataset}. Note that for contact PPG signal interbeat intervals are estimated from diastolic minima since they are more clear and prominent than peaks.} 
	
 	\label{fig:PPGsignal}
\end{figure}

\item Another approach is to assess average pulse rate as frequency corresponding to maximal power spectral density (PSD). 
By computing PSD over $N$ points one estimates average pulse rate value over time interval $\tau = N/F_\text{SR}$, where $F_\text{SR}$ is the sampling rate of the iPPG signal ($F_\text{SR} = 50$~Hz for DEAP). 
PSD is usually estimated by {\bf Discrete Fourier Transform (DFT)} or by {\bf autoregressive (AR) modeling}. 

{\bf DFT} is a direct way to estimate PSD \citep{PohMcDuffPicard2011, deHaanJeanne2013}. 
Yet, DFT is often criticized \citep{HuelsbuschBlazek2002,HoltonEtAl2013} since its frequency resolution is $60/\tau$~BPM ($1/\tau$~Hz) and leads to a crude estimation of pulse rate for $\tau < 20$~s, while taking  $\tau > 20$~s hinders tracking of pulse rate variations. 
Here we use $N = 1024$, which results in averaging pulse rate over $\tau = 20.48$~s.
	
{\bf AR modeling} considers iPPG as an output of linear system with added white noise \citep{TakanoOhta2007, Tarassenko2014AR}; 
	parameters of this system are estimated to compute PSD.  
	In comparison with DFT, AR modeling yields improved resolution for short samples. 
	We implement AR modeling using Burg's method (Matlab function \texttt{pburg}) and employ models either of $23$-rd order (for iPPG signal with wavelet filtering at Step 4) or $34$-th order (without wavelet filtering) as these settings provide best pulse rate estimation (we have tested orders $5, \ldots, 80$). 
	
\item {\bf Continuous Wavelet Transform (CWT)} provides a promising alternative to DFT and AR modeling  \citep{HuelsbuschBlazek2002}.
We implement CWT using Matlab function \texttt{cwtft}; we take Morlet wavelet \citep{Huelsbusch2008, BousefsafMaaouiPruski2013} and scales\footnote{We choose these scales to have a sufficiently good coverage of human heart rate bandwidth (0.65--4Hz): 0.325~Hz is twice lower than minimal pulse rate, while 25~Hz is the half of iPPG signal sampling rate. Factor $2^{0.03125}$ provides 32 scales per octave.} corresponding to 0.325 -- 25~Hz with factor $2^{0.03125}$.  
\end{itemize}

Since DFT and AR modeling estimate only average pulse rate, in order to make all estimates comparable, we average pulse rate estimates for IBI and CWT in windows of $\tau = 20.48$~s.

Note that methods of pulse rate estimation have been recently compared in \citep{CuiFuHongZhangShu2015} for iPPG extracted using G method, but in that study CWT and AR modeling were not considered, while DFT was used either with long windows of 30~s resulting in low time resolution or with short windows of 2~s providing very low frequency resolution.





\subsection{Metrics}\label{subsec:Metrics}
To investigate quality of pulse rate estimation, we split each trial (see Section~\ref{subsec:Dataset}) into epochs of 20.48~s with 9.88~s (approximately 50\%) overlap and get five epochs per trial. 
For each epoch $i$ we compare estimated average pulse rate $\text{PR}_i$ with the averaged reference value $\text{PR}^\text{ref}_i$. 
The following quantities are used to assess estimation performance for the epochs of each participant. 

{ \bf Mean absolute error (MAE)} is given by
\begin{equation}
	\text{MAE} = \frac{1}{N} \sum_i \vert \text{PR}_i - \text{PR}^\text{ref}_i \vert, 
\end{equation}
where $N = 5 \text{ epochs per trial} \times \text{amount of trials}$ and $i$ is the number of epoch. 
$\text{MAE}\approx3$~BPM was observed in \citep{Tarassenko2014AR} for epochs comprising 4 heart beats (approximately 4~s) and $\text{MAE}\approx2.5$~BPM on average in \citep{LewandowskaRuminskiKocejkoNowak2011} for 30~s epochs. 
		
{ \bf  Root-mean-square error (RMSE)} is given by
\begin{equation}
	\text{RMSE} = \frac{1}{N} \sqrt{\sum_i \big( \text{PR}_i - \text{PR}^\text{ref}_i \big)^2}. 
\end{equation}		
$\text{RMSE}$ is more sensitive to large estimation errors than $\text{MAE}$, so small number of large errors results in high $\text{RMSE}$ and low $\text{MAE}$. 
Pulse rate estimates from uncompressed video of stationary subjects usually have RMSE in range of 1--2 BPM for epochs of 30 -- 60 s \citep{PohMcDuffPicard2011, LiChenZhaoPietikainen2014, BousefsafMaaouiPruski2016}\footnote{$\text{RMSE} < 1$~BPM observed in \citep{deHaanJeanne2013} is obtained for video recorded under dedicated professional illumination, which makes results incomparable with those for DEAP. On the other hand, $\text{RMSE} > 7.6 BPM$ reported in \citep{LiChenZhaoPietikainen2014} for MAHNOB dataset is too high and indicates limited usefulness of this dataset for iPPG-based pulse rate estimation.}. 
		
		
{ \bf Percentage of epochs (PE)} for those pulse rate is estimated with error below $3.5$~BPM\footnote{In \citep{HoltonEtAl2013} best method estimates pulse rate with error below $6$ BPM for $\text{PE}_6 = 87$\% of epochs. 
Here we are interested in percentage of epochs for those pulse rate is estimated well; precision of $6$ BPM seems insufficient for this, so we bound error by $3.5$~BPM (5\% of average human pulse rate 70 BPM).} is given by
\begin{equation}
	\text{PE}_{3.5} = \frac{1}{N} \lbrace i : \vert \text{PR}_i - \text{PR}^\text{ref}_i \vert < 3.5 \text{ BPM} \rbrace. 
\end{equation}					 		


We also assess quality of iPPG signal by {\bf signal-to-noise ratio (SNR)} defined as \citep{deHaanJeanne2013}:
\begin{equation}
\text{SNR} = \frac{1}{N} \sum_i 10\log_{10}\frac{\sum\limits_{f=40\text{ BPM}}^{240\text{ BPM}} \big(\hat{S}_i(f)\big)^2U_i(f)}
							   {\sum\limits_{f=40\text{ BPM}}^{240\text{ BPM}} \big(\hat{S}_i(f)\big)^2\big(1-U_i(f)\big)}, 
\end{equation}
where $\hat{S}_i(f)$ is the spectrum of the $i$-th iPPG epoch computed by using DFT and $U_i(f)$ indicates whether
frequency component $f$ is attributed to the signal ($U_i(f) = 1$) or to noise ($U_i(f) = 0$):
\begin{equation*}
U_i(f) = \begin{cases}
			1,&	\text{if } \vert f - \text{PR}^\text{ref}_i \vert \leq \Delta f \text{ or if } \vert f - 2\text{PR}^\text{ref}_i \vert \leq 2\Delta f,\\ 			
			0,&	 \text{otherwise}. 			
		 \end{cases}	
\end{equation*}
In order to make results comparable with those in \citep{WangDenBrinkerStuijkDeHaan2017} we take 
$\Delta f = \frac{50\cdot60}{1024} \approx 2.93$~BPM.

\section{Results and Discussion}\label{sec:Results}
\subsection{Overview}\label{subsec:MainResult}
In Table~\ref{tab:MeanAverageError} we present quality metrics fo pulse rate estimates and iPPG extraction methods under best pre- and post-processing (MCaS detrending is beneficial for all methods, other pre- and post-processing methods providing best results are summarized in Table~\ref{tab:PrePostProcessing}). 
In all cases best results are obtained for whole face ROI with skin selection and outliers rejection (see Figure~\ref{fig:testScheme},  Step 1).  

\begin{table}[h] 
\caption{Quality metrics (averaged over all epochs and participants) for pulse rate estimates computed from iPPG with best pre- and post-processing (Table~\ref{tab:PrePostProcessing}). Values in each cell stand for MAE (BPM) / RMSE (BPM)/ $\text{PE}_{3.5}$ (\%); best values of metrics for each iPPG extraction method are shown in {\bf bold}.} 
\label{tab:MeanAverageError}

 \centering
 \begin{tabular}{ l | c | c |  c | c  }
 \toprule
iPPG 		&\multicolumn{4}{c}{Pulse rate estimation}\\
extraction 	& IBI				& DFT				& AR						& CWT\\
\midrule
G 		  	& 5.91 / 7.42 / 44	& 6.54 / 9.62 / 54	& 5.62 / {\bf 6.78} / 44	& {\bf 5.35} / 7.62		  / {\bf 58}\\
GRD 	 	& 4.05 / 5.30 / 60	& 3.82 / 6.36 / 73 	& 3.99 / 	5.11\, 	/ 60	& {\bf 3.07} / {\bf 4.96} / {\bf 78}\\ 
aGRD 	 	& 4.41 / 5.70 / 57	& 4.41 / 7.07 / 70 	& 4.28 / {\bf 5.45}	/ 58	& {\bf 3.59} / 5.55		  / {\bf 74}\\ 
ICA 	 	& 3.91 / 5.41 / 64	& 3.61 / 6.03 / 75 	& 3.71 / 	4.87\, 	/ 63	& {\bf 2.94} / {\bf 4.77} / {\bf 79}\\ 
CHROM 	 	& 3.46 / 4.64 / 65	& 2.70 / 4.70 / 81 	& 3.05 / 	4.04\, 	/ 69	& {\bf 2.08} / {\bf 3.46} / {\bf 86}\\ 
POS 	 	& 3.13 / 4.30 / 70	& 2.61 / 4.50 / 81 	& 2.91 / 	3.80\, 	/ 71	& {\bf 1.99} / {\bf 3.25} / {\bf 87}\\  
\bottomrule
 \end{tabular}
 \end{table}

The lowest estimation errors are achieved when using POS for iPPG extraction and CWT for pulse rate estimation. 
Altogether, values of quality metrics are comparable with those reported in the literature for pulse rate estimation from uncompressed video (see Subsection~\ref{subsec:Metrics}). 
 
\begin{table}[htbp] 
\caption{Pre- and post-processing providing most precise pulse rate estimates (processing for IBI and DFT also results in highest SNR); 
IIR and FIR stand for corresponding band-pass filtering, MA$x$ -- for moving average filtering with length $x$, WF -- for wavelet filtering. 
} 
\label{tab:PrePostProcessing}

 \centering
 \begin{tabular}{ l | c | c | c |  c | c }
 \toprule
iPPG 		&\multicolumn{2}{c|}{Pre-processing}&\multicolumn{3}{c}{Post-processing}\\
extraction 	&IBI, DFT, CWT 	& AR				& IBI, DFT	&  CWT 			& AR\\
\midrule
G 		  	&\multicolumn{2}{c|}{--}			&\multicolumn{3}{c}{MA12, FIR, WF}\\
GRD 	 	&\multicolumn{2}{c|}{--}			&\multicolumn{2}{c|}{MA12, FIR, WF}	&MA12, IIR, WF\\
aGRD 	 	&FIR					&IIR		&\multicolumn{3}{c}{MA12, WF}\\
ICA 	 	&FIR, MA12				&IIR, MA12	&\multicolumn{3}{c}{WF}\\
CHROM 	 	&\multicolumn{2}{c|}{--}			&MA12, FIR, WF	& MA9, FIR ,WF		&MA12, WF\\
POS 	 	&\multicolumn{2}{c|}{--}			&MA12, FIR, WF	& MA9, FIR			&MA12, WF\\
\bottomrule
 \end{tabular}
 \end{table}

Below we discuss influence of various steps on the pulse rate estimation quality. 
We begin with methods for iPPG extraction (see Figure~\ref{fig:testScheme}, Step 3), since results for different methods vary considerably (Subsection~\ref{subsec:ExtractionResult}). We proceed with ROI selection (Step 1), pre- and post-processings (Steps 2, 4) and finish with pulse rate estimation (Step 5) in Subsections~\ref{subsec:ROIresult}--\ref{subsec:PulseResult}, respectively.

\subsection{Step 3: Imaging photoplethysmogram extraction}\label{subsec:ExtractionResult}
In Table~\ref{tab:ResultExtraction} we present performance metrics for all considered iPPG extraction methods.   
We use pre- and post-processing (Steps 2, 4) ensuing best SNR (see Table~\ref{tab:PrePostProcessing}) and estimate pulse rate by CWT since this method provides best results at Step 5. 

\begin{table}[htbp] 
\caption{Average SNR and quality metrics for CWT pulse rate estimates from iPPG extracted using pre- and post-processing settings providing best SNR. 
In each cell we present values with and without wavelet filtering. 
For comparison we include 
overall SNR values from \citep{WangDenBrinkerStuijkDeHaan2017}.
Best values for each quality metrics are shown in {\bf bold}.
} 
\label{tab:ResultExtraction}

 \centering
 \begin{tabular}{ l | c | c |  c | c  | c  }
 \toprule
iPPG extraction & $\text{MAE}$, BPM	& $\text{RMSE}$, BPM&$\text{PE}_{3.5}$, \%	&$\text{SNR}$, dB& reported SNR, dB\\
\midrule	
G 		 		& 5.40 / 5.60		& 7.70 / 8.20		& 58 / 58				& -3.24 / -4.16	&-1.90	\\ 
GRD 			& 3.15 / 3.11		& 5.00 / 5.08		& 77 / 78 				& -0.89 / -2.14	&\;3.67 \\ 
aGRD 		 	& 3.59 / 3.63		& 5.55 / 5.75		& 74 / 74				& -1.26 / -2.45	&was not considered \\  
ICA 			& 2.98 / 3.02		& 4.88 / 5.11		& 79 / 80 				& -0.60 / -1.93	&\;1.92	\\ 
CHROM 		 	& 2.10 / 2.12		& 3.39 / 3.56		& 85 / 86				& -0.20 / -1.67	&\;3.86	\\ 
POS 			&2.04 / {\bf 2.01}	&{\bf 3.16} / 3.17	& {\bf 87 / 87}			& {\bf 0.30} / -1.19	&\;{\bf 5.16}	\\ 
\bottomrule
 \end{tabular}
\end{table}

As you can see from Table~\ref{tab:ResultExtraction}, POS has the highest signal-to-noise ratio and provides most precise pulse rate estimation. 
The ranking of methods is generally in line with results reported in \citep{WangDenBrinkerStuijkDeHaan2017}, except for GRD performing worse than ICA. 
We explain this difference by sensitivity of ICA to the number of source components in the signal; light variation  and motion in \citep{WangDenBrinkerStuijkDeHaan2017} introduce additional components to the color signals and may complicate extraction of pure iPPG by means of ICA. 

Note that average SNR for DEAP dataset is worse than values reported in \citep{WangDenBrinkerStuijkDeHaan2017}. It might be due to the compression of videos in DEAP and using professional dedicated lighting for video acquisition in \citep{WangDenBrinkerStuijkDeHaan2017}. 

\begin{figure}[!htbp]
	\centering
	\includegraphics[scale=0.55]{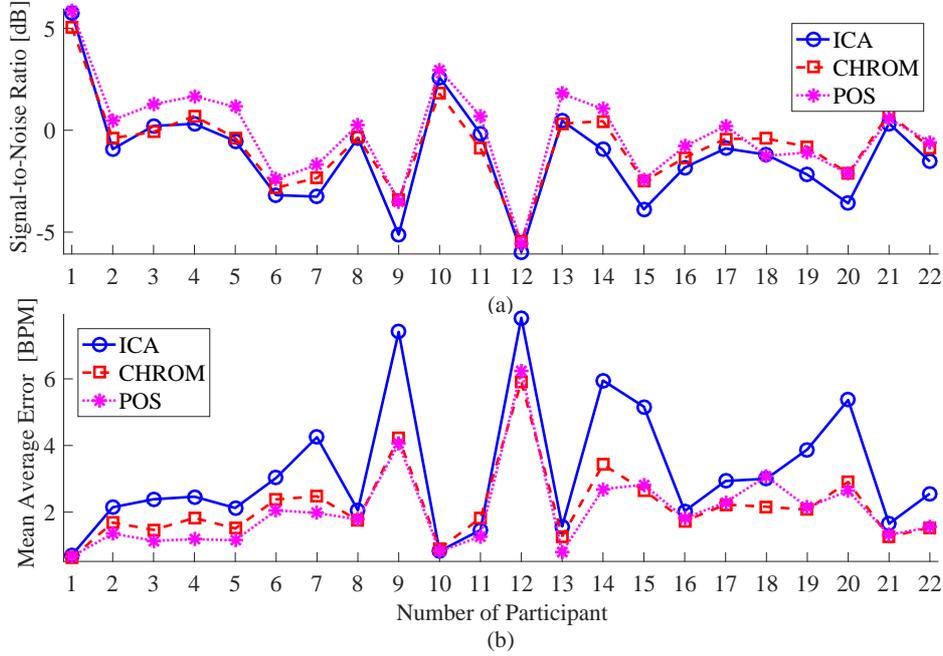}
	\caption{$\text{SNR}$ (a) and $\text{MAE}$ (b) obtained by ICA, CHROM and POS methods for iPPG extraction in combination with best pre- and post-processing settings (Table~\ref{tab:PrePostProcessing}) and CWT pulse rate estimation.}
 	\label{fig:ResultParticipantwise}
\end{figure}

Figure~\ref{fig:ResultParticipantwise} shows average values of $\text{SNR}$ and $\text{MAE}$ for every participant for three best iPPG extraction methods (POS, CHROM and ICA). 
In most cases high $\text{SNR}$ corresponds to low $\text{MAE}$, which (as expected) indicates that good quality of iPPG ensures precise pulse rate estimation.  

\subsection{Step 1: ROI Selection}\label{subsec:ROIresult}
{\bf ROI choice.} Results for the whole face region are better than for the region below eyes, both in terms of iPPG quality and pulse rate estimation. Namely, SNR for signal acquired from the whole face region is at least 0.2-0.3 dB higher (for GRD and aGRD; for other methods the difference is 1.1-1.9 dB), while MAE of pulse rate estimation is lower (contribution varies from 1\% for GRD and aGRD to more than 10\% for other methods).	

{\bf ROI refinement.} Outliers rejection is always beneficial, as it increases SNR of iPPG signal (0.25--0.5~dB) and decreases MAE (10\%--15\% when CWT or DFT pulse rate estimation is used in combination with GRD, aGRD, CHROM or POS methods and 5\%--10\% otherwise). 

\subsection{Steps 2 and 4: Pre-processing and Post-processing}\label{subsec:ProcessingResult}
{\bf Detrending.} Among two detrending methods, MCaS always improves pulse rate estimation while SPA does not provide any positive effect (probably raw color signals in our study are too noisy for successful application of this technique).  
For G, aGRD and ICA methods of iPPG extraction (Step 2), using MCaS increases SNR (average increase is 1.6, 0.7 and 3 dB, respectively) and improves pulse rate estimation (decrease of MAE is above 12\%). 
For CHROM, POS and GRD methods using MCaS at pre-processing is immanent, therefore performance without MCaS was not tested.  

\begin{figure}[!htbp]
	\centering
	\includegraphics[scale=0.55]{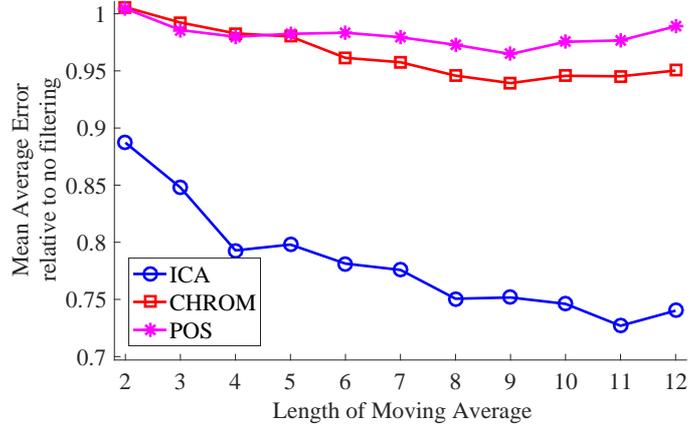}
	\caption{Average $\text{MAE}$ values reflecting the influence of $M$-point MA filtering on pulse rate estimation using CWT (pre- and post-processing according to Table~\ref{tab:PrePostProcessing}).
}
 	\label{fig:MAeffect}
\end{figure}

{\bf MA filtering}. Quality of iPPG signal and of pulse rate estimation enhances with increase of MA filter length $M$ and reaches maximum for $M=12$ (MA filtering with $M>12$ affects heart rate bandwidth and was not tested). 
The only exception is pulse rate estimation by CWT (Step 5) from iPPG obtained by CHROM and POS methods: in this case best results are observed for $M = 9$.  
Figure~\ref{fig:MAeffect} illustrates this effect for MAE, effect on other quality metrics is similar. 

{\bf Band-pass filtering} improves quality of iPPG signal and, in most cases, performance of pulse rate estimation, see Table~\ref{tab:bandPassEffect}. Surprisingly, band-pass filtering has little positive or even negative effect on pulse rate estimation by AR modeling
; we cannot explain this result. 
The 255th order FIR filter performs slightly better than the 5th order IIR Butterworth filter; this was expected since frequency response of the latter is slightly worse. 
\begin{table}[htbp] 
\caption{Effect of band-pass filtering on SNR and MAE estimated from iPPG with pre- and post-processing according to Table~\ref{tab:PrePostProcessing}. In each cell the first value is for the 5th order IIR Butterworth filter and the second for the 255th order FIR filter.} 
\label{tab:bandPassEffect}

 \centering
 \begin{tabular}{ l | c | c |  c | c | c  }
 \toprule
iPPG 		& \multicolumn{4}{c|}{decrease of MAE relative to the value without band-pass filtering, \%}	&SNR increase, \\
extraction 	&\;\;\;\;\;IBI\;\;\;\;\;&\;\;\;\;\;DFT\;\;\;\;\;&\;\;\;\;\;AR\;\;\;\;\;	& CWT 	&	dB\\
\midrule
G 		  	& 25 / 21 				& 14 / 20				& 10 / 10 			& 22 / 31	& 0.38 / 0.66\\
GRD 	  	& 22 / 13 				& 13 / 20				&  6 /  5			& 22 / 31	& 0.29 / 0.55\\
aGRD 	 	& 22 / 17 				& 15 / 19				&  7 /  6 			& 21 / 29	& 0.31 / 0.57\\
ICA 		& 19 / 30 				&  7 / 11				&  3 /  0 			& 14 / 21 	& 0.22 / 0.42\\
CHROM 	 	& 13 / 26  				&  6 / 11				& -1 / -6 			& 11 / 16	& 0.19 / 0.38\\ 
POS 	  	& 15 / 27 				&  9 / 14				& -1 / -6 			& 14 / 20	& 0.21 / 0.39\\  
\bottomrule
 \end{tabular}
 \end{table}

{\bf Wavelet filtering} results in elimination of large errors in pulse rate estimation which is reflected by prominent decrease of RMSE (see Table~\ref{tab:ResultExtraction}). 
However, wavelet filtering only slightly improves MAE and almost does not change $\text{PE}_{3.5}$. 
Parameter choice for the wavelet filter deserves a separate study: 
preserving several harmonics of pulse rate is of interest to keep the shape of iPPG signal. 

{\bf Pre- vs post- processing.} 
Band-pass and MA filtering are preferable at pre-processing (Step 2) when ICA or aGRD are used at Step 3 and at post-processing (Step 4) for other methods of iPPG extraction, see Table~\ref{tab:PrePostProcessing}. 
For ICA, POS and CHROM using filtering at a different step considerably decreases quality of iPPG and of pulse rate estimation.
This is quite unexpected since originally POS and CHROM were proposed with band-pass filtering as pre-processing \citep{deHaanJeanne2013, WangDenBrinkerStuijkDeHaan2017}, while for ICA post-processing was recommended \citep{McduffGontarekPicard2014OCG}. 
For aGRD band-pass filter is essentially a pre-processing sub-step and we do not observe any difference between using MA filter as pre- and post-processing. 


\subsection{Step 5: Pulse Rate Estimation}\label{subsec:PulseResult}
\begin{figure}[!htbp]
	\centering
	\includegraphics[scale=0.55]{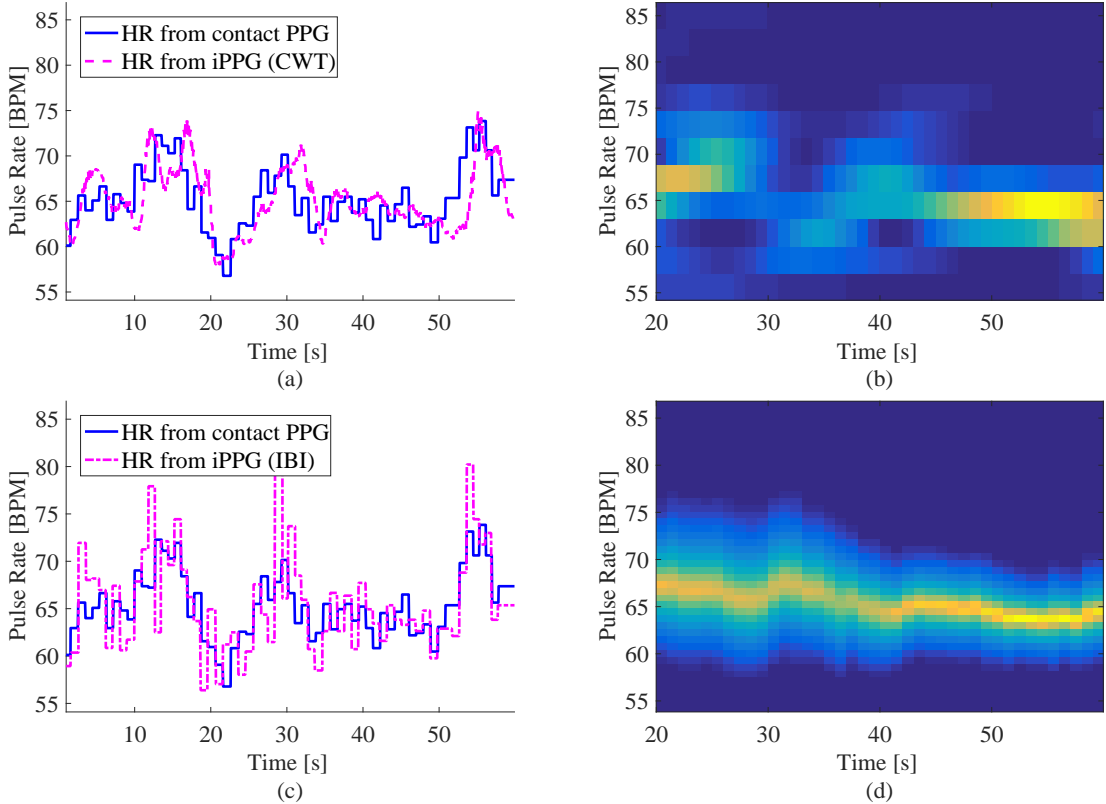}
	\caption{Performance of pulse rate estimation methods for iPPG signal extracted using POS method from data P1 T24: momentary pulse rate estimated by CWT (a, smoothed by 1-s moving average) and IBI (c), spectrograms of iPPG signal estimated using DFT (b) and by AR modeling (d).}
 	\label{fig:VariousPulseEstim}
\end{figure}

The best results in terms of all metrics are provided by {\bf CWT}.
This method is especially useful since it allows to estimate not only average but also momentary pulse rate (see Figure~\ref{fig:VariousPulseEstim}a).

Other tested methods for pulse rate estimation have certain drawbacks. 
{\bf DFT} provides the second best result in terms of MAE and $\text{PE}_{3.5}$ (Table~\ref{tab:MeanAverageError}), but it has low frequency resolution (see Figure~\ref{fig:VariousPulseEstim}b) and highest RMSE.  
{\bf IBI estimation} (Figure~\ref{fig:VariousPulseEstim}c) has the lowest overall performance that can be explained by the insufficient quality of iPPG extracted from compressed  DEAP videos. 
IBI filtering techniques \citep{McduffGontarekPicard2014OCG} may improve precision of IBI estimation. 
Finally, {\bf AR modeling} requires an elaborate choice of model order as using orders different from those selected in Subsection~\ref{subsec:PulseRate} resulted in considerably worse pulse rate estimation. 


\section{Conclusion}\label{sec:Conclusion}
Let us summarize the main results of this work. 
We have established a generic framework for iPPG-based pulse rate estimation. 
Using this framework we have compared various methods of iPPG analysis for compressed video from DEAP dataset; best pulse rate estimation is obtained when using following methods.
\begin{description}
	\item [Step 1, ROI selection:] whole face ROI with skin selection and outliers rejection.  		  
	\item [Step 2, pre-processing:] mean-centering and scaling; moving average filtering (for ICA) with filter length $M$ close to $\frac{1}{4}F_\text{SR}$, where $F_\text{SR}$ is sampling rate in Hz; band-pass filtering (for ICA and aGRD) with 255th FIR filter.  	
	\item [Step 3, iPPG extraction:] POS; result for CHROM and ICA are also relatively good.
	\item [Step 4, post-processing:] moving average and band-pass filtering (if not used at pre-processing), wavelet filtering.
	\item [Step 5, pulse rate estimation:] Continuous Wavelet Transform. 
\end{description}


Let us finish with two problems that may become interesting topics for the further research.
\begin{itemize}
	\item Here we have considered only pulse rate estimation, but one can also use DEAP dataset to investigate estimation of pulse rate variability \citep{McduffGontarekPicard2014remote} and respiratory rate \citep{Tarassenko2014AR} from iPPG.  
	\item Up to now iPPG has been used only for human subjects. However, iPPG acquisition does not seem unfeasible for non-human animals with bare face, for instance, for primates \citep{ChangiziZhangShimojo2006bareSkin}. Using iPPG for pulse rate estimation can be beneficial for animal research, where using contact measurements is often undesirable.   
	 
\end{itemize} 

\subsection*{Acknowledgements}
The author acknowledges funding from the Ministry for Science and Education of Lower Saxony and the Volkswagen Foundation through the program “Niedersächsisches Vorab”. Additional support was provided by the Leibniz Association through funding for the Leibniz ScienceCampus Primate Cognition.

The author is grateful to MPI for Dynamics and Self-Organization for the support during the work at the manuscript.
The author thanks Dr. S.~M{\"o}ller, Dr. I.~Kagan and Prof. F.~Wolf for useful discussions and remarks on the manuscript.   


\end{document}